\documentclass[prd,floats,preprint,superscriptaddress,eqsecnum,floatfix,nofootinbib,12pt]{revtex4}

\usepackage{amsmath}
\usepackage{graphics, graphicx}
\usepackage{epsfig}
\usepackage{ulem}
\usepackage{color}
\usepackage{mathtools}

\def\ah{a}
\def\Nh{N}
\def\phih{\phi}
\def\cS{{\cal  S}}

\def\vx{{\vec x}}
\def\scrip{{\cal I}^+}
\def\1p{{(1p)}}
\def\vx{{\vec x}}
\def\hij{h_{ij}}
\def\scrim{{\cal I}^-}

\def\td{TD\ }
\def\p0{\phi_0}
\def\pnb{p_{NB}}
\def\ptra{p_{\rm trans}(\p0'',\p0')}
\def\ptr{p_{\rm trans}}
\def\scrim{{\cal I}^-}
\def\be{\begin{equation}}
\def\ee{\end{equation}}
\def\beq{\begin{eqnarray}}
\def\eeq{\end{eqnarray}}
\def\qf{}
\def\rf{}

\def\zf{}
\def\jf{}
\def\j2{}
\def\mf{}
\def\rf{}
\def\kf{}
\def\tf{}
\def\uf{}
\def\nf{}
\def\p0{\phi_0}
\def\td{t_d}
\def\z0{\zeta_0}

\def\cV{{\cal V}}

\def\epb{{(2)}}

\newcommand{\ttle}[1]{{\it #1}}

\begin{document}

\title{Quantum Transitions Between Classical Histories}

\author{James  Hartle}
\affiliation{Department of Physics, University of California, Santa Barbara,  93106, USA}
\author{Thomas Hertog}
\affiliation{Institute for Theoretical Physics, KU Leuven, 3001 Leuven, Belgium}

\bibliographystyle{unsrt}

\begin{abstract}
In a quantum theory of gravity spacetime behaves classically when quantum probabilities are high for histories of geometry and field  that are correlated in time by the Einstein equation. Probabilities follow from the quantum state. This quantum perspective on classicality has important implications:
(a) Classical histories are generally available only in limited patches of the configuration space on which the state lives. (b) In a given patch, states generally predict relative probabilities for an ensemble of possible classical histories. (c) In between patches classical predictability breaks down and is replaced by quantum evolution connecting classical histories in different patches. (d) Classical predictability can break down on scales well below the Planck scale, and with no breakdown in the classical equations of motion. 
We support and illustrate (a)-(d) by calculating the quantum transition across the de Sitter like throat connecting asymptotically classical, inflating histories in the no-boundary quantum state. This supplies probabilities for how a classical history on one side transitions and branches into a range of classical histories on the opposite side. We also comment on the implications of (a)-(d) for {\rf the dynamics of black holes and eternal inflation}.

\end{abstract}

\vspace{1cm}


\maketitle
\section{Introduction}
\label{intro}

A quantum system behaves classically when the probability is high for suitably coarse-grained histories of its motion that exhibit correlations in time governed by deterministic, classical, dynamical laws. The classical behavior of the flight of a tennis ball, the orbit of the Earth, and the spacetime of the universe are all examples. This paper is concerned with the classical behavior of spacetime {and in particular with its limitations in the context of quantum cosmology}. 

Probabilities for the individual members of (decoherent) sets of alternative coarse-grained histories of a closed system like the universe follow from the system's quantum state and dynamics. The state can be represented by a wave function $(\Psi)$ on an appropriate configuration space. The dynamics is summarized by an action $(I)$ defined on histories which are curves in that space. 

{ Probabilities predicting
classical evolution} may not be available for all times and in all regions of the configuration space on which the wave function $\Psi$ is defined.  A wave packet scattering off a barrier in one dimension provides a simple example. Suppose the center of the wave packet follows a classical trajectory before it hits the barrier. After clearing the barrier the wave function will have branched into transmitted and reflected wave packets which behave classically. But, during the time it is interacting with the barrier, there will generally be no classical behavior at all. Classical behavior has broken down during that time, although there is no breakdown in the classical equations. However the two domains of classical behavior can be connected quantum mechanically with the Schr\"odinger equation as we show in more detail in Section \ref{barrpen}. 

{ In the semiclassical approximation to quantum cosmology}, the quantum state of a closed universe $\Psi$  is specified by a wave function on the superspace of three-geometries $h_{ij}(\vx)$  and matter field configurations $\chi(\vx)$ on a closed spacelike three-surface $\Sigma$. Schematically we write $\Psi=\Psi[h,\chi]$.  In this paper we will argue on the basis of various examples that
\begin{enumerate}
 
\item[(a)] Classical histories of geometry and field are generally available only in limited patches of superspace. 
 
 \item[(b)] In a given patch states generally do not predict one classical history, but rather an ensemble of possible ones with probabilities for the individual histories in the ensemble.   
 
 \item [(c)] In between patches classical {evolution} breaks down and is replaced by quantum evolution that supplies probabilities for connecting a given classical history in a given patch with any {\j2 classical} history in another patch. One history {\jf typically} branches into many histories. 
 
 \item[(d)] Classical {\zf evolution} based on quantum probabilities can break down on scales {\jf much larger  than} the Planck length, and with no breakdown in the classical equations of motion. 
 
 \end{enumerate}
\noindent Therefore:
 \begin{itemize}  
 
 \item  Global concepts and results of classical general relativity will generally not be appropriate in physical situations where classical spacetimes exist only in patches of superspace. In particular notions such as absolute event horizons, domains of dependence, singularity theorems, etc may not be applicable. 
 
 \item  In general there will be no evolution of states of a field (unitary or otherwise) through a foliating family of spacelike surfaces in a classical background spacetime where no such background exists. 

 \end{itemize}   
 
{\jf This quantum perspective on classicality can have significant implications} for discussions of black hole evaporation{\j2 and eternal inflation}. We will speculate on this in Sections \ref{blackholes} and \ref{ei} returning in more detail to them in  future papers. Here we concentrate mainly on the example of quantum transitions in the bouncing universes predicted by the no-boundary wave function (NBWF) \cite{NBWF}. This illustrates all of the above features in an elementary way. For this analysis we build on a series of papers \cite{HHH08a,HHH08b,HHH10a,HHH10b,HH13} devoted to calculating and investigating the classical universes predicted by the NBWF. 

The paper is organized as follows: Section \ref{model} discusses the minisuperspace model and the regions of classical behavior in minisuperspace (MSS). Section \ref{qtransitions} discusses a few simple examples of quantum transitions. In Section \ref{thrubounce} and \ref{probhist} the quantum mechanical transition amplitudes between two sides of a cosmological bounce are approximated. Possible implications of a quantum perspective on classicality for evaporating black holes and for the dynamics associated with eternal inflation are discussed in Section \ref{blackholes} and\ref{ei} {\j2 respectively.}  Section \ref{conclusions} has some brief conclusions. 

\section{Classical Prediction in the No-Boundary Quantum State}
\label{model}

\subsection{Histories of the Universe --- Classical and Quantum}
\label{histories}

We consider a minisuperspace model in which the Lorentzian four-geometries are homogeneous, isotropic, and spatially closed on the manifold $M={\bf R}\times S^3$. {\tf We take a positive cosmological constant $\Lambda \equiv 3H^2$ and a single homogeneous scalar field $\Phi$ moving in a quadratic potential $V=(1/2) m^2\Phi^2$ as a model of the matter content of our universe.} This provides a simple {\tf context} for illustrating the points (a)-(d) in the introduction\footnote{\jf We should stress that the model is not intended to be a realistic description of the universe. Realistically there are fluctuations away from homogeneity and isotropy. We treat these somewhat in Sections \ref{flucts} and \ref{ei}.}.

The metrics of homogeneous, isotropic, Lorentzian (quantum mechanical) histories in standard coordinates are 
\begin{subequations} 
\begin{align}
ds^2 &= -\Nh^2(\lambda) d\lambda^2 + \ah^2(\lambda) d\Omega^2_3  \\
           &=-dt^2  +\ah^2(t) d\Omega^2_3 
\label{homoiso}
\end{align}
\end{subequations}
where $d\Omega^2_3$ is the metric on the unit round 3-sphere, $\ah$ is the scale factor, and $dt=Nd\lambda$.  For convenience we will work with a rescaled field $\phi\equiv(4\pi/3)^{1/2}\Phi$. More details on our conventions can be found in Appendix \ref{appa}.

Quantum states are represented by wave functions $\Psi$ on the configuration space spanned by the three-geometries and the field configurations on a spacelike surface $\Sigma$. This configuration space will be called {\it superspace} in this paper. Taking $\Sigma$ to be a surface of homogeneity in \eqref{homoiso}, useful coordinates for minisuperspace are the scale factor of the three-geometry which we denote by $b$ and the homogeneous value of the scalar field $\phi$ denoted\footnote{As in our previous papers we find it convenient to maintain a notational distinction between the variables $(b,\chi)$ spanning the configuration space of the wave function and the variables $(a,\phi)$ which describe four-dimensional histories of geometry and field. But of course, histories of the form \eqref{homoiso} are specified by functions $(\ah(\lambda)$, $\phih(\lambda))$ that define curves in configuration space $(b(\lambda),\chi(\lambda))$ and vice versa.} by $\chi$. Thus $\Psi=\Psi(b,\chi)$. When there is a need to be more compact we write ${x^A} =(b,\chi), \  A=1,2$. Then $\Psi=\Psi(x^A)$ or even $\Psi=\Psi(x)$. 

{\it Classical histories} of the form \eqref{homoiso} are specified by functions $\ah(\lambda)$ and $\phih(\lambda)$ that satisfy the classical Einstein equation and the dynamical equations for the scalar field \eqref{loreqns}. Classical histories are predicted in regions of superspace where the wave function is well approximated by  {\mf a sum of terms of} semiclassical (``WKB'') form
\be
\Psi(x^A) \approx A(x^A) \exp[i S(x^A)/\hbar]
\label{semiclassical-wf}
\ee
with $S$ varying rapidly compared to $A$ over the region. That is roughly when
\be 
\label{classcond}
|\nabla_A S| \gg |\nabla_A A/A|,  \quad \text{(Classicality Condition)}.
\ee
When this classicality condition is satisfied, a straightforward generalization\footnote{This generalization can be derived from quantum mechanics, see, e.g. \cite{Har95c}.} of WKB leads to the prediction of an ensemble of classical histories that are the integral curves of $S$. That is, they are the solutions of the equations 
\be
p_A = \nabla_A S
\label{momenta}
\ee
where $p_A$ are the momenta {\jf containing time derivatives that are conjugate to the $x^A$.} {\mf Each term defines a } classical history through each point in configuration space with momentum determined by \eqref{momenta}. The space of classical histories {\tf determined by $\Psi$} is thus half the dimension of the phase space of the theory. The probability $p$ of the history that passes through $x^A$ is, to leading order in $\hbar$, given by 
\be
\label{probhist1}
p \propto |A(x^A)|^2 .
\ee
This is conserved along the the history a consequence of the operator implementation of the Hamiltonian constraint --- the Wheeler-DeWitt equation \cite{HHH08b}. 

This semiclassical algorithm for classical prediction has been extensively used in quantum cosmology to extract predictions for cosmological observables from a wave function of the universe\footnote{In the framework of decoherent (or consistent) histories quantum mechanics in which we are working probabilities can be assigned to a set of alternative histories only when they decohere, that is, when there is negligible quantum interference between the individual members in the set (see, e.g. \cite{Har93a} for a tutorial). In many realistic situations histories following one variable are decohered by interactions with other variables often called an environment. However our simple minisuperspace model does not contain enough variables to model such mechanisms. We are therefore assuming that when decoherence of spacetime can be realistically modeled the probabilities will not be signficantly different from those we calculate here.}. We now briefly summarise the classical predictions of the no-boundary wave function (NBWF) \cite{NBWF} in the above minisuperspace model.

\subsection{Classical Histories {\uf of}  the Semiclassical NBWF}
\label{nbwf-classens}

We have discussed the ensemble of homogeneous isotropic classical histories predicted by the NBWF in considerable detail in \cite{HHH08a,HHH08b}. Here we offer a brief summary of those results. 

The NBWF $\Psi[h,\chi]$ is formally given by an integral over histories of geometry $g$ and fields $\phi$ on a four-{disk} with one boundary $\Sigma$. The contributing histories match the values $(h(x),\chi(x))$ on $\Sigma$ and are otherwise regular. They are weighted by $\exp(-I/\hbar)$ where $I[g,\phi]$ is the Euclidean action \eqref{curvact} - \eqref{mattact} for gravity coupled to the scalar field. The sum is taken along a complex contour chosen so that the defining integral converges and the result is real {\tf so that the NBWF is time neutral}. Implemented for the homogeneous isotropic minisuperspace model defined in Section \ref{model} the sum defining $\Psi(b,\chi)$ is over those $N(\lambda)$, $b(\lambda)$,  and $\phi(\lambda)$ that define regular geometries and fields on the disk and match $(b,\chi)$ on the boundary.  

In some regions of superspace the path integral can be approximated by the method of steepest descents. There the NBWF will be approximately given by a sum of terms of the form 
\begin{equation}
\Psi(b,\chi) \approx  \exp[-I(b,\chi)/\hbar] = \exp\{[-I_R(b,\chi) +i S(b,\chi)]/\hbar\} ,
\label{semiclass}
\end{equation}
one term for each complex extremum of the action that is regular on the four-disk and matches the real values $(b,\chi)$ on its boundary.  (See \eqref{euceqns_N} for the {\uf equations} that determine these extrema.) The functions $I_R(b,\chi)$ and $-S(b,\chi)$ are the real and imaginary parts {\tf respectively}  of the Euclidean action  evaluated at the extremum, $I(b,\chi)$.  Eq \eqref{semiclass} has the semiclassical form \eqref{semiclassical-wf}. Hence, in regions of superspace where $S$ varies rapidly compared to $I_R$, we predict a one parameter ensemble of classical histories, one through each point $(b,\chi)$ in the region that  the approximation holds. As in \eqref{probhist1},  the NBWF probability for the history passing through $(b,\chi)$ is proportional to $\exp[-2I_R(b,\chi)/\hbar]$. 

The one parameter ensemble of classical histories could be labeled by the value of $\chi$ at some fiducial value of $b$.  But for consistency with our earlier work we prefer to label the histories by $\p0$ which is the absolute value of the scalar field at the south pole of the complex saddle point to which the classical history corresponds \cite{HHH08a}. {\zf A remarkable property of the classical ensemble predicted by the NBWF is that 
{\it all} classical histories have an early period of scalar field driven inflation \cite{HHH08b,Lyo92}. During inflation the classicality conditions \eqref{classcond} hold and, in the $\hbar=c=G=1$ Planck units that we use throughout,  these solutions take the {\j2 approximate} form
\begin{subequations}
\label{slowroll}
\begin{align}
\chi(t) & = \p0 -mt/3 , \\
b(t)&= \frac{1}{2m\p0}\exp\left(m\p0 t -\frac{1}{6}m^2 t^2\right) .
\label{slwroll}
\end{align}
\end{subequations}
This is just a subset of the usual family of inflationary solutions to the classical slow roll equations for a quadratic potential.
The individual histories are labeled by $\p0$ whose range we take to be bounded above by the Planck scale at $\p0\sim 1/m$ {\tf beyond which our methods of calculation probably no longer apply}. {\j2 The NBWF probability that the history labeled by $\p0$ occurs is
\be
\label{nbprob}
p_{NB}(\p0) \propto \exp[-2I_R(\p0)/\hbar]
\ee
where $I_R(\p0)$ is the real part  of the action of the saddle point corresponding to $\p0$. }

The number of efolds $N$ is  $N \sim \p0^2$. {\zf Figure \ref{classhist} shows a few examples of different types of classical histories in the ensemble for this minisuperspace model when the ratio $\mu \equiv (3/\Lambda)^{1/2}m >3/2$. The trajectories shown in this figure were obtained by naively extrapolating the classical histories \eqref{slowroll} both to the future and into the past of the slow roll era using the classical equations \eqref{loreqns}. {\nf (We will see below that this approximation is not always justified!)} For values $\p0 \gg {\cal O}(1)$ the classically extrapolated histories bounce at a minimum radius $b_m \approx  (m \p0)^{-1}$, never reaching a singularity \cite{HHH08a}. By contrast for $\p0 \sim {\cal O}(1)$ the classical histories begin with a singularity at infinite scalar field and zero scale factor, and either expand forever or recollapse to another singularity\footnote{ \tf If $\mu>3/2$. There turn out to be  no classical histories with $0<\p0 \lesssim 1$ \cite{HHH08b}.}.

\begin{figure}[t]
\includegraphics[height=3.2in]{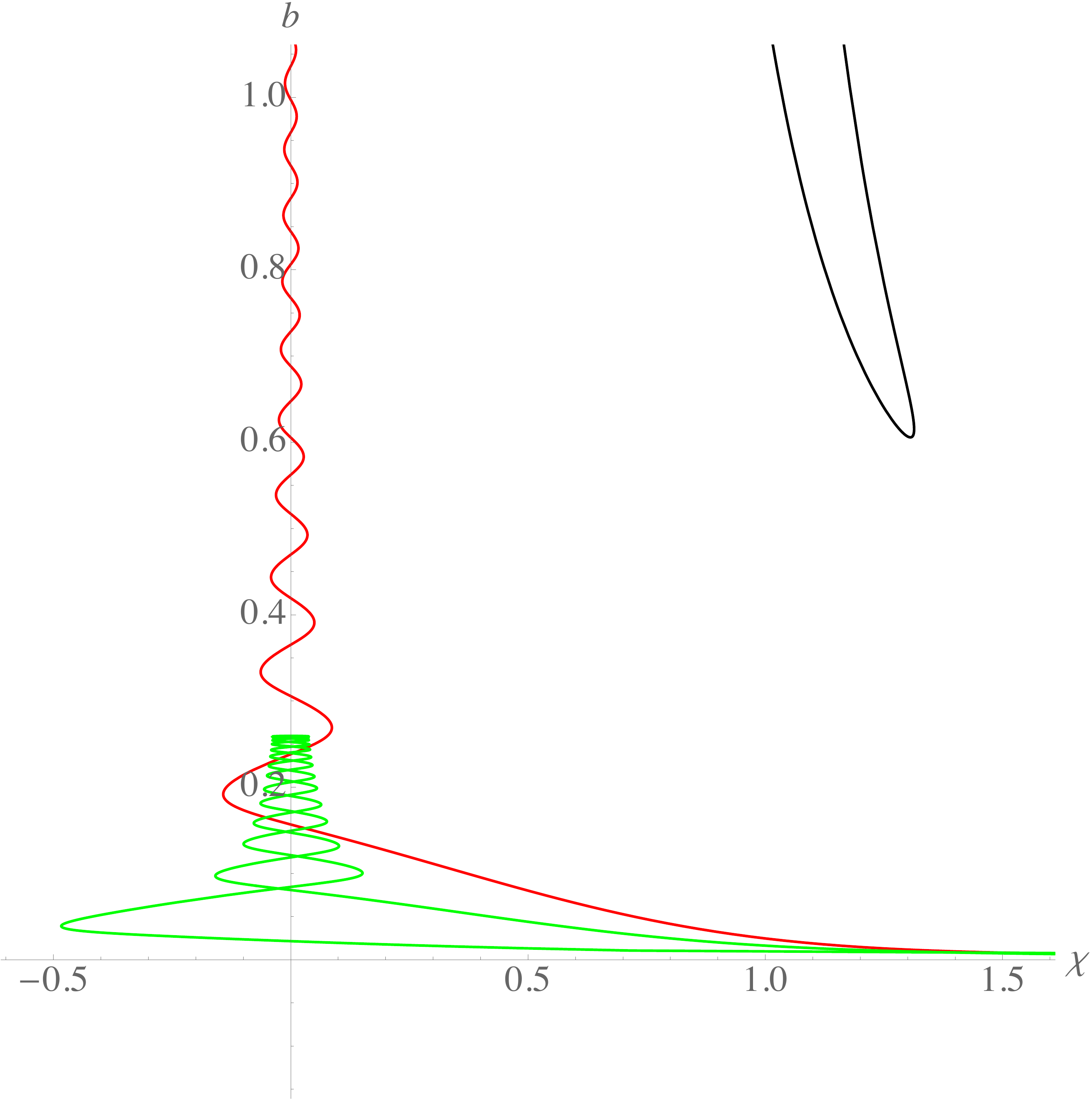} 
\caption{A history is a curve in minisuperspace not necessarily single-valued in any variable. Classical histories are particular curves satisfying the classical equations of motion \eqref{loreqns}. Three kinds of classical cosmological histories are shown: Bouncing histories that contract to a minimum radius $b_m$ and then re-expand, never reaching a singularity (black curve). Histories that start at a singularity at infinite scalar field $\chi$ and zero $b$ can expand forever (red) or recollapse to another singularity (green).} 
\label{classhist}
\end{figure}

The NBWF is real. This means that for every saddle point of the form \eqref{semiclass} there is also a complex conjugate saddle point with $S$ replaced by $-S$ {\tf giving rise {\uf to a second ensemble} of classical histories.} Reversing the sign of $S$ is the same as reversing the direction of time [cf. \eqref{momenta}]. For every classical history in the first ensemble, therefore, its time reversed is in the second {\uf with the same probability \eqref{nbprob}.} If we think of one set of histories as expanding, in the other set they are contracting. {\zf The classical ensemble predicted by the real NBWF consists of the union of both sets}. As we will show in Section \ref{thrubounce}, it is possible to think of a classical history in one ensemble as `connected' by a quantum transition to a classical history in the other ensemble to make a complete {\it quasi classical} history of a bouncing universe.

\subsection{The Region of Validity of the Semiclassical Approximation}
\label{regsc}

Classical histories follow from the semiclassical NBWF only in regions of superspace where $S$ varies rapidly compared to $I_R$ -- the classicality condition  \eqref{classcond}.

{\zf {\jf To approximate the region where the classicality conditions hold we use a simple analytic approximation to the  Euclidean action of the complex saddle points associated with the classical histories \eqref{slowroll}.} The saddle points have a representation in which a slightly deformed four-sphere of radius $(m \p0)^{-1}$ is joined smoothly onto a Lorentzian inflationary spacetime of the form \eqref{slowroll}. The Euclidean four-sphere region approximately determines $I_R$ whereas the Lorentzian regime provides the main contribution to $S$, yielding}
\be
\label{lyons-acts}
I_R(b,\chi) \approx - \frac{\pi}{(2m\chi)^2},  \quad  S(b,\chi) \approx \frac{m \chi b^3}{3}
\ee
From this \cite{HHH08b} one easily calculates 
\be
\label{lyo-cc}
\left|\frac{\nabla_\chi I_R}{\nabla_\chi S}\right| \approx \frac {1}{(m\chi b)^3}, \quad
\frac{(\nabla I_R)^2}{(\nabla S)^2} \approx  \frac{1}{(m b \chi)^6 \chi^2 } .
\ee
Here $\nabla_A$ denotes the superspace derivative and inner products are in the superspace metric \eqref{GV}.  Requiring the left hand sides of  \eqref{lyo-cc} to be much less than unity requires, for $\chi\gtrsim 1$, that 
\be
\label{sc-bound}
b \gtrsim \frac{1}{m\chi}
\ee
and a slightly stronger inequality for $\chi \lesssim 1$. 

\begin{figure}[t]
\includegraphics[height=3.6in]{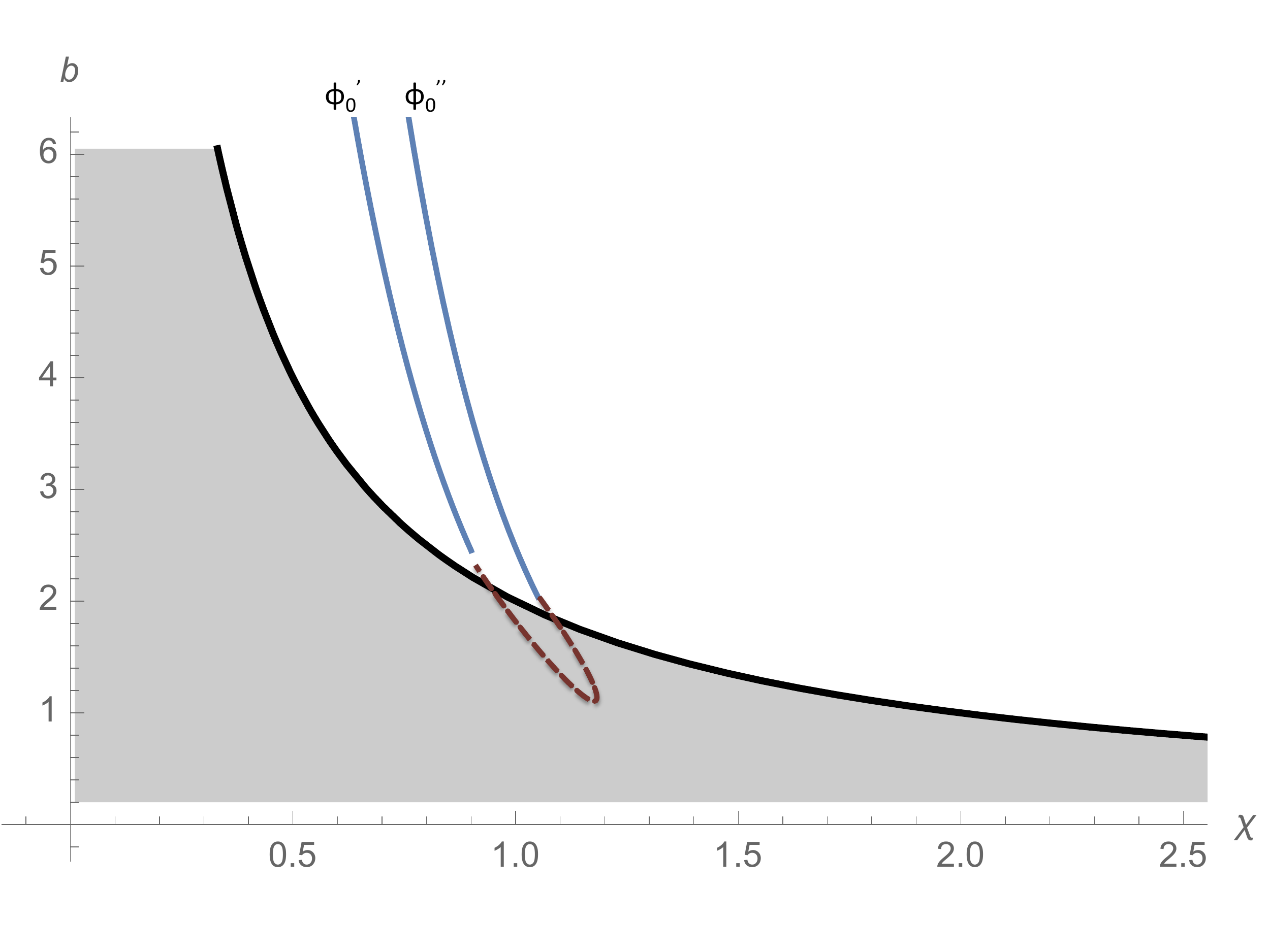} 
\caption{Classical prediction in minisuperspace.   We expect no classical behavior when the field is above the value $\chi_{\rm pl}\sim1/m$ corresponding to the Planck scale field energy density. The classical equations of motion are well defined over the whole of the region  $b>0$, $\chi<\chi_{\rm pl}$. But in this region the NBWF probabilities only predict classical behavior in the unshaded patch where the classicality condition holds according to \eqref{sc-bound}. In that region two representative classical histories $b=b(\chi)$ are shown. These are labeled by $\p0$ which increases from low $\chi$ to high $\chi$ . There are quantum transitions between these histories as described in Section \ref{thrubounce}. }
\label{scbound}
\end{figure}

Since each history has an early inflationary phase driven by an effective cosmological constant with
\be
H_{\rm eff}(\p0) \equiv m\p0
\label{effH}
\ee
we can rewrite the condition for classicality \eqref{sc-bound} in the following intriguing way
\be
S_{dS} \lesssim A(b).
\label{areabnd}
\ee
Here, $S_{dS} \equiv \pi /H_{\rm eff}^2$ is the {\zf effective} deSitter entropy \cite{GH77} and $A(b) \equiv 4\pi b^2$ is the area of a maximal two-surface in a spatial section. In this form the condition for classicality in the NBWF ensemble is similar to the holographic entropy bound
\cite{Bous02}. Since the NBWF can be formulated holographically \cite{HH11b} it is not entirely out of the question that this is more than a numerical coincidence. 

{\zf Figure \ref{scbound} shows the region of superspace where the classicality condition holds according to \eqref{sc-bound} along with two representative classical histories in the large $\p0$ domain. When extrapolated classically each history would bounce at a minimum radius $b_m \approx  (m \p0)^{-1}$, as discussed above. However we now see that a classical extrapolation of the histories through the bounce is not justified in quantum cosmology, because the wave function does not predict classical behavior in the regime where the bounce occurs.}

\subsection{The Breakdown of Classicality, the Planck Scale, and Classical Equations}

Eq \eqref{sc-bound} shows that classicality breaks down when the radius of the universe becomes less than
\be
\label{classbound}
b \sim \frac{1}{m\chi} .
\ee
In the Planck units used in this paper the potential energy $V=(1/2)m^2\chi^2$ reaches the Planck energy $\sim 1$ when the scalar field is $\chi\sim (1/m)$. The classicality bound on the radius \eqref{sc-bound} is $1$ ---one Planck length. The semiclassical approximation plausibly breaks down when Planck energies are reached. {\zf But \eqref{classbound} shows that the classicality condition also breaks down for smaller values of $\chi$ well below the Planck scale. Hence there is a regime of radii much larger than the Planck length where classical evolution does not  emerge from the NBWF.}

{\j2 Next consider the classical equations of motion.} The Lorentzian equations of motion \eqref{loreqns} and their solutions remain well behaved in the region where the classicality condition is not satisfied. Indeed, in \cite{HHH08a} we computed the trajectories of the bouncing histories (cf. Fig.\ref{classhist}) using the classical equations of motion. That is, we extrapolated a classical history from the region of superspasce where the classicality holds into the small $b \sim b_m$ region where it breaks down without any problems. As we said there 
``We will assume that once classical histories have been identified in a region of minisuperspace where the classicality condition holds they can be extrapolated to regions where it does not hold using the classical equations of motion until they become classically singular. That is an assumption which can in principle be checked in the full quantum mechanical theory.'' This paper can be regarded as making that check, {\zf and eq. \eqref{sc-bound} shows it is far from obvious that classical extrapolation through a bounce is a good approximation to the quantum mechanical evolution. {\jf In fact, a classical extrapolation through the bounce is not precisely invariant under time reversal \cite{HHH08a}. This is already an indication that there should be some quantum corrections because, as discussed above, the ensemble of classical histories predicted by the NBWF consists of two copies of the same ensemble.}

In a quantum universe classical predictability is a matter of high quantum probabilities for histories with deterministic correlations in time. The lack of classical predictability is {\mf not a matter of the classical equations of motion}.  The magnitude of quantum probabilities cannot be reliably diagnosed  from the classical equations of motion  for cosmology any more than it can for any other quantum system. It requires a quantum mechanical analysis from the system's quantum state like the semiclassical NBWF in this paper, such as given in Section \ref{regsc}. 

We emphasize that histories of $(a,\phi)$ do not end simply because a semiclassical approximation breaks down. Histories are defined on the whole of the manifold $M={\bf R} \times S^3$. A history need not be all quantum or all classical. A history may be classical in some regions of configuration space but not generally where the classicality condition is not satisfied. What a history does when not classical is a matter of quantum mechanical probabilities supplied by the wave function $\Psi$. {\zf In the next Section we will show that these probabilities allow histories to make {\it probabilistic quantum transitions} through the region of semiclassical breakdown between different histories in the classical region of superspace. In this way we are led to a quantum version of a bounce. }

\section{Quantum Transitions --- Examples}
\label{qtransitions}

{\tf To drive home the point that classical predictability can break down even when the classical equations of motion do not,  we consider two familiar elementary examples --- barrier penetration and the cosmological amplification of quantum fluctuations. These are analogs of quantum transitions across bounces in cosmology in this respect but not in all. They differ {\uf for example} in the physical mechanisms that give rise to the quantum corrections.}

\subsection{Barrier Penetration}
\label{barrpen}

Barrier penetration in non-relativistic quantum mechanics provides a very simple model of ensembles of classical histories connected by quantum transitions. Consider a single particle of mass $M$ moving in one-dimension ($x$)  in a potential $V(x)$ that is a barrier at around $x=0$ and zero far away). Suppose that at $t=t_0$ the state is a wave packet $\psi_{(x_0,p_0)}(x, t_0)$  centered around a large negative position $x_0$ and a positive momentum $p_0$ with widths consistent with the uncertainty principle. Assume that all times of interest are small enough that wave packet spreading can be neglected.

 The center of the wave packet initially moves on a classical trajectory
\be
\label{classtraj}
x(t) = x_0 + p_0 (t-t_0)/M.
\ee
Coarse grain position at one time by a set of intervals $\Delta_\alpha$, $\alpha = 1,2, \cdots$ with each interval wider than the width of the wave packet.  A history of position is given by a sequence of intervals $\alpha=(\alpha_1, \alpha_2, \cdots \alpha_n)$ at a series of times $t_1,\cdots, t_n$. Up until the time $t_{b} $ that the wave packet first interacts the barrier this state will predict one classical history following \eqref{classtraj} with near unit probability.  By a later time $t'_{b}$ the wave packet will have split into a reflected wave packet and a transmitted wave packet outside the barrier each centered around a classical trajectory like \eqref{classtraj} but with different values of $x_0$ and $p_0$. The integrals of the squares of the wave packets give the probabilities that the particle is reflected or transmitted. 

Thus before $t_b$ there is a classical ensemble consisting of one classical history which has probability $1$.
After $t'_b$ there is a classical ensemble consisting of two classical histories with probabilities that are the reflection and transmission probabilities for the barrier.   In between these two times the particle does not behave classically. To connect the two classical ensembles we solve the Schr\"odinger equation or, equivalently, calculate the propagator,
\be
\label{nrprop}
\langle x', t'_b | x, t_b \rangle = \int_{[x t_b, x' t'_b]} \delta x \exp [i {\cal S}[x(t)]/\hbar] ,
\ee
where the path integral is over particle paths that connect $x$ at $t_b$ to $x'$ at $t'_b$. The propagator defines a quantum transition matrix between{\j2 histories}  of the classical ensembles before $t_b$ and after $t'_b$.  The one initial classical history before is connected probabilistically to the two histories afterwards. 

{\qf By  way of example, consider the transition between the initial wave packet $\psi_{(x_0,p_0)}(x, t_0)$ at $t_0<t_b$ described above and a second similar wave packet $\psi_{(x_1,p_1)}(x, t_1)$ at $t_1>t'_b$. The transition amplitude between these states  is 
\be
\label{transbarrier}
T_{(x_0,p_0)\rightarrow(x_1,p_1)}=\int dx' \int dx  \ \psi^*_{(x_1,p_1)}(x', t_1)\langle x', t_1 | x, t_0 \rangle\psi_{(x_0,p_0)}(x, t_0) .
\ee
This can be regarded as the transition amplitude between the  two classical histories labeled by $(x_0,p_0)$ at  $t_0$ and $(x_1,p_1)$ at  $t_1$. The transition probability is the square of this amplitude. If the wave packets are sufficiently narrow in $x$ this will be proportional to $|\langle x_1, t_1 | x_0, t_0 \rangle|^2$. 

{\qf The four points (a)-(d) stressed in the Introduction are explicitly modeled by barrier penetration{\j2 as follows:}

 \begin{enumerate}
 
 \item[(a)] For the given initial  state classical histories of the particle are available only in the two patches of $(t,x)$-space, $t<t_b$ and $t>t'_b$.
 
 \item[(b)] In the patch $t<t_b$ there is a classical ensemble with  only one coarse-grained classical history having probability 1. In the patch $t>t'_b$ the classical ensemble consists of two classical histories with probabilities that the particle is reflected or transmitted. One {\j2 history} goes into many.
 
 \item [(c)] In between patches classical predictability breaks down and is replaced by quantum evolution summarized by \eqref{nrprop} with gives probabilities for transitions between the classical histories in the two patches. 
 
 \item[(d)] The classical equation of motion $Mdx/dt=-dV/dx$ does not break down{\j2 anywhere anytime}. It remains well defined but it  gives an incorrect answer --- either reflection or transmission depending on the initial momentum. 
 
  \end{enumerate}
 }

\subsection{Cosmological Fluctuations}
\label{flucts}

The evolution of linear fluctuations about a homogeneous, isotropic background cosmological geometry provides another {\jf example that nicely illustrates the implications of a quantum perspective on classicality.}
We merely sketch the argument here referring the reader to \cite{HHH10a} for details.

To describe the quantum evolution of perturbations away from closed, homogeneous and isotropic three-geometries and field configurations we consider a minisuperspace model consisting of the scale factor $b$, the homogeneous value of the scalar field $\chi$, and parameters $z=(z_1,z_2,...)$ specifying  perturbation modes. Thus, $\Psi=\Psi(b,\chi, z)$. We use the NBWF as a model state. {\jf Its semiclassical approximation is given by \eqref{semiclass} but now with an action that includes the fluctuation degrees of freedom. Expanding the saddle point action in the fluctuations gives 
\begin{equation}
I(b,\chi,z) = I^{(0)}(b,\chi) + I^\epb (b,\chi,z) .
\label{pertaction}
\end{equation}
where $I^\epb$ is quadratic in $z$'s. 
From this and \eqref{semiclass} we can identify a wave function for the fluctuations in the semiclassical approximation
\be
\psi(z;b,\chi) \propto \exp[-I^\epb (b,\chi,z)/\hbar] .
\label{semiclass-flucts}
\ee
When evaluated on one of the homogeneous isotropic classical histories $(b(t),\chi(t))$ predicted by the background wave function, we get a time dependent wave function for the fluctuations evolving in this background\footnote{This can be more carefully derived from the defining functional integral for the NBWF, see \cite{HHH10a}.}.
\begin{equation}
\psi(z,t) \equiv \psi(b(t),\chi(t),z).
\label{wfot}
\end{equation} 
As shown in a variety of ways (see e.g. \cite{Har87}),  the Wheeler-DeWitt equation then implies a Schr\"odinger equation for $\psi(z,t)$
\begin{equation}
i\hbar \frac{\partial\psi(z,t)}{\partial t} = H(t) \psi(z,t)  . 
\label{seqn}
\end{equation}
The time dependent Hamiltonian describes the evolution of the state of the fluctuations in the background $(b(t),\chi(t))$. In this way, the fluctuation fields can be thought of as quantum fields on the possible background classical spacetimes.  We have recovered quantum field theory in curved spacetime. 

Classical histories of fluctuations are predicted in patches of fluctuation minisuperspace where the classicality conditions are satisfied for the action $I^{(2)}(b,\chi,z)$. For a given mode $z$ this occurs approximately at the time when it leaves the horizon, that is, when the wavelength of the mode exceeds the Hubble radius at that time. 

Thus initially in $t$ we have an ensemble of homogeneous and isotropic classical histories as before, now augmented by non-classical fluctuations $z$ described by a quantum wave function $\psi(t,z)$. As time progresses the homogenous isotropic classical geometries branch into an ensemble of perturbed geometries with different classical fluctuations.  This illustrates the observations (a)-(d) in the Introduction as follows:

 \begin{enumerate}
 
 \item[(a)] For a particular fluctuation mode a classical history is available in a limited region of superspace at a sufficiently late time after the mode has left the horizon.
 
 \item[(b)] The classical ensemble consists of homogeneous, isotropic backgrounds with many modes of classical fluctuations. 
 
 \item [(c)] Before the time when a mode becomes classical, its evolution is described by the Schr\"odinger equation \eqref{seqn}. 
 
 \item[(d)] The linear classical equations for fluctuations do not break down. It is only that they are not implied by quantum mechanical probabilities until the mode leaves the horizon.
 
  \end{enumerate}

}

\section{Quantum Bounces}
\label{thrubounce}

\subsection{Quasiclassical Histories of the Universe}
\label{qchistories}

{ In Section \ref{nbwf-classens} we reviewed how the NBWF in the large volume regime predicts two identical sets of classical, inflationary, asymptotically de Sitter histories. One set of histories is expanding, the other set is the time-reversed and thus describes contracting universes. But we have also seen (cf. Section \ref{regsc}) that classical evolution must break down in the no-boundary state at small volume, when the size of the universe is comparable to the scale set by the effective cosmological constant. In particular we showed in \cite{HHH08a} that the expanding and contracting histories in the classical NBWF ensembles are not connected by classical evolution across a bounce, even when the classical extrapolation describes a bouncing history where the curvature is everywhere low, and the classical equations remain everywhere well defined. That is, the classical extrapolation across a bounce of a history in the first set does not yield a member of the second set. The asymptotic time-symmetry\footnote{Individual histories need not be time-symmetric, but their time reverse must also be a member of the ensemble with the same probability. Hence the time-neutrality of the no-boundary quantum state implies time symmetry in a statistical sense.} implied by the NBWF therefore leads to quantum corrections to the classical evolution in the region of the bounce\footnote{Because the NBWF predicts the same classical ensemble on both ends it can also be thought of as imposing both initial and final quantum states in a time neutral formulation of quantum mechanics that allows these \cite{GH93b,HHXX,Har95c}. This has an effect on the evolution near the bounce.}.

In this section we estimate the quantum transition between the expanding and contracting branches of the NBWF. The resulting histories will be {\it quasiclassical} histories on the manifold ${\bf R}\times S^3$ that describe a phase in which the universe contracts classically from large to small volumes and executes a quantum transition through a ``bounce'' to (a range of) classically expanding NBWF histories on the far side.}

\subsection{Parametrized Proper Time}
\label{paramtime}

{\tf To carry out the construction of a quantum connection between the two NBWF ensembles of classical histories we introduce a `ideal clock' variable $t$} so that $t\rightarrow \pm\infty$ labels the two asymptotic ensembles on opposite sides of the bounce\footnote{We sometimes refer to `initial' negative values of $t$ and `final' positive ones. But as we shall see the ensemble of predicted histories is time symmetric.}. A standard way of doing this is to define a proper time coordinate by $dt=Nd\lambda$ in \eqref{homoiso} and promote that to the status of a canonical variable with a conjugate momentum $p_t$  \cite{Partime}.  This process, called `parametrizing the proper time', is described in Appendix \ref{paramproper}.  The resulting theory is $t-$reparametrization invariant with a consequent constraint of the form
\be
p_t + H(p_A,x^A)=0\ .
\label{pconstraint}
\ee
Here $x^A=(b,\chi)$, $p_A$ are the conjugate momenta, and $H(p_A,x^A)$ is the usual Hamiltonian constraint for these MSS models \eqref{Hconstr}.  Thus we are in an extended minisuperspace MSSt spanned by $(x^A,t)$.
The quantum state now becomes a function of $t$ as well as of the $x^A$, $\Psi=\Psi(x^A, t)$ .   
 Expressing the momenta in \eqref{pconstraint} as operators and fixing an operator ordering  we find that the operator form of the constraint becomes the Schr\"odinger equation
\be
i\hbar\frac{\partial\Psi}{\partial t} = H\left(-i\hbar\frac{\partial}{\partial x^A}, x^A\right)\Psi .
\label{pschrod}
\ee
 The NBWF can be thought of as a particular time independent solution of this equation. 

The time $t$ can be thought of as a matter degree of freedom that keeps track of proper time and has a Hamiltonian that is linear in the momentum with a spectrum ranging over the whole real line\footnote{As is well known \cite{Pauli58}, negative energies must be allowed to have a matter variable keep exact track of the time in the Schr\"odinger equation.}. The explicit form of this equation for our extended minisuperspace model is given in \eqref{seqn-explicit}. We now use it to provide a quantum connection between the classical histories in the ensembles at large positive and negative values of $t$. 

Introducing a new degree of freedom is a departure from the action used to calculate the classical histories of the NBWF.  However, we intend to use it only to estimate the transition probabilities between classical histories in the two NBWF ensembles. We will give arguments below that introducing this clock variable should yield reasonable estimates of these probabilities\footnote{An alternative that does not introduce any other degrees of freedom would be to investigate the evolution of wave packets in MSS as in \cite{Kiefer}. {\tf It would be of interest to explore this connection.}}. 

\subsection{Asymptotic Classicality} 
\label{asympclass}
The Schr\"odinger equation \eqref{pschrod} implies classicality at large values of $b$ at asymptotic values of $t$. To show this we calculate the semiclassical (WKB) approximation to its stationary states
\be
\Psi \propto \exp{[i(W(x,E) -Et)/\hbar] } .
\label{scstates}
\ee
Here,  to leading order in $\hbar$, $W(x,E)$ obeys the classical Hamilton-Jacobi equation
\be
\label{HJ1}
H\left(\frac{\partial W}{\partial x^A} ,x^A\right)= E\ .
\ee
The specific form of this equation for our MSS model is given in \eqref{HJ-explicit}. For the semiclassical approximation to the NBWF set  $E=0$ and then $S(b,\chi)=W(b,\chi)$.

For large values of $b$ the solutions of the Hamilton-Jacobi equation \eqref{HJ1} can be expanded in powers of $b$. We investigated this expansion \cite{Starobinsky83} for essentially this MSS model in \cite{HH12b,HHH14}. The first three leading terms are universal. (We need only the leading term). For large $b$ and small values of $\chi$ we found
\be
\label{asympc}
W(b,\chi) = \frac{H}{3} b^3 + \frac{H}{2}\lambda_{-} \chi^2 b^3 +  {\cal O}(b^2)
\ee
where $\lambda_{\pm} \equiv (3/2)(1\pm\sqrt{1-4m^2/(9H^2)}$ and we assume ranges of $m$ and $H$ where this is real\footnote{Realistic values of $m$ and $H$ are not in this range but we aim at a tractable model\footnotemark[1].}. The effect of the constant $E$ shows up in lower order terms in the expansion of order $\log(b)$. Therefore it has a relatively small effect on the asymptotic behavior. 
Therefore at large $b$ the action $W$ in \eqref{scstates} is varying rapidly so these states will each predict an ensemble of asymptotic classical histories that are solutions of \cite{HH12b,HHH14}
\be 
\frac{db}{dt} = \frac{1}{b} \frac{\partial W}{\partial b},   \quad \frac{d\chi}{dt}=-\frac{1}{b^3} \frac{\partial W}{\partial\chi} \ .
\label{HJeqns}
\ee
Thus, in this MSS model, the Schr\"odinger equation \eqref{pschrod} naturally implies ensembles of classical histories at large times $\pm t$ when $b$ is large. For the NBWF we recover in this way the asymptotic form of the subset of classical histories \eqref{slowroll} in the NBWF ensembles which continue to expand as $t \rightarrow \pm \infty$. Asymptotically the expansion is not driven by the scalar field potential energy but by the cosmological constant. Hence the solutions of \eqref{HJeqns} are asymptotic deSitter spaces with different histories of a decaying scalar field. 

We now use the Schr\"odinger equation \eqref{pschrod} to estimate the quantum transition probabilities between members of both classical ensembles at large negative and positive $t$.  See Figure \ref{bnc-hist}. 
\begin{figure}[t]
\includegraphics[width=4.in]{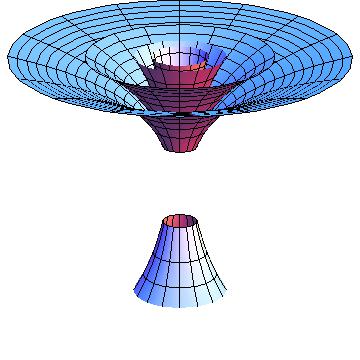}
\caption{ An embedding diagram that schematically shows a single  initial classical contracting universe (bottom) that makes a quantum transition to any one of an ensemble of final classical expanding universes (top) with different probabilities.  The universe contracts to a size given by \eqref{classbound} and approximately reemerges at the same size  after a quantum transition in the intervening region. A slice of this figure is shown in Fig \ref{transition}. One can think of parametrized proper time $t$  as running from the bottom of the figure to the top.} 
\label{bnc-hist}
\end{figure}

\begin{figure}[t]
\includegraphics[width=4.in]{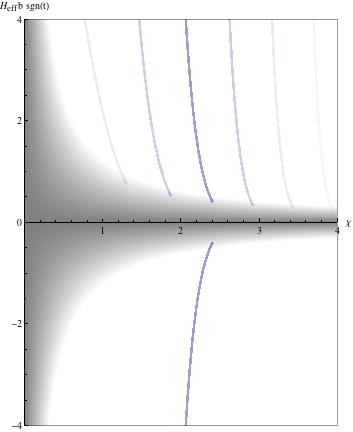}
\caption{This figure shows a single classical history (bottom)  labeled by $\p0'$ making a quantum transition to an ensemble of classical histories on the other side of a bounce (top).{\j2 The vertical axis is essentially the parametrized time $t$ but labeled more usefully by $H_{\rm eff}b$  multiplied by the sign of $t$. $H_{\rm eff}\equiv m\p0$ is the effective cosmological constant near the bounce. The horizontal axis  is the scalar field $\chi$.}{\j2 The classicality condition \eqref{classcond} is not satisfied in the shaded region and classical histories are not predicted there. The curves in the top half of the figure are members of one classical ensemble and the one in the bottom half belongs to the time reversed ensemble. }The single history at the bottom  contracts until it reaches the shaded region defined approximately by \eqref{classbound} where the classicality condition is not satisfied. It then makes a quantum transition to one of a final ensemble of expanding histories (top)  distinguished by values of  $\p0''$.  The transition is mediated by the propagator \eqref{propagator} which is a sum over all the histories connecting points on the initial and final histories --- classical or not. An estimate of the transition probabilities is \eqref{transprob-ex}. The largest probability is for a transition to the time-symmetric expanding history (as suggested by the shading).} 
\label{transition}
\end{figure}

\subsection{Transition Amplitudes}
\label{transamps}

The quantum transition amplitude between two classical histories is specified by the Schr\"odinger equation \eqref{pschrod}, or equivalently, by the propagator between two points in MSSt,
\be
\label{propagator}
K( x'',  t'' | x' , t') \equiv \int_{[x't',x''t'']} \delta x  \exp{\{i {\cal S}_p [x(t)]/\hbar\}} .
\ee
Here, $x =(b,\chi)$ labels a MSS point  and ${\cal S}_p$ is the action associated with the Hamiltonian in \eqref{pschrod} i.e. the gauge fixed, time-parametrized action for Lorentzian MSSt histories \eqref{rpa-fixed}. The sum is over all paths $x^A(t)$  in  MSSt  that are single valued\footnote{This simplifying assumption is in fact not an assumption, see \cite{Har95c} Sec 7.3.} in $t$  and  connect $(x',t')$ to $(x'',t'')$. 
Interchanging $(x',t')$ with $(x'',t'')$ reverses the sign of the action in \eqref{propagator} with the result that 
\be
\label{timerev}
K( x'',  t'' | x' , t') = K^*( x',  t' | x'' , t'') \ .
\ee
{\j2 We hope that the analogy with barrier penetration in Section \ref{barrpen} is evident and supportive. The constraint equation \eqref{pschrod} is analogous to the Schr\"odinger equation, and the propagator \eqref{propagator} is analogous to \eqref{nrprop}.}

{\mf To estimate the transition amplitude $T(\p0'',\p0')$ between two classical histories labeled by $\p0'$ and $\p0''$ we pick points on the different histories and evaluate the propagator \eqref{propagator} between them.  To get a rough estimate  it is convenient to pick these points in regimes of superspace where (i) the classicality condition \eqref{classbound} is satisfied, and (ii) in the slow roll regime of each history where \eqref{slowroll} is valid. (Then intersections between histories can be ignored.) We assume that this region is large enough that the transition probabilities stabilize as the points are moved to larger times along the histories in this regime thus defining a transition amplitude between histories and not just between points\footnote{Greater accuracy could possibly be obtained by tracking wave packets following each history (cf. \cite{Kiefer}). }.}

We thus obtain a transition matrix between classical histories that in many ways is analogous to an S-matrix.  Since physically we cannot specify the points to arbitrary accuracy we smear the propagator with functions $e^{\sigma}_{\p0}(x)$ that extend over a width $\sigma^{-1}$ around the point $x$ on history $\p0$ with the value of $\sigma$ large compared to the scale on which $\p0$ varies significantly. This can also be thought of as part of the coarse-graining necessary for classicality. The result is [cf.\eqref{transbarrier}] 
\be
\label{transamp}
T(\p0'',\p0') \propto  \lim_{t',t'' {\rm large}} \int dx'' \int dx' e^{\sigma}_{\p0''}(x'') K(x'',t''|x',t')e^{\sigma}_{\p0'}(x') 
\ee
assuming the limits as above. 
The transition probabilities between classical histories are related by
\be
\label{transprob}
p_{\rm trans}(\p0'',\p0') \propto |T(\p0'',\p0')|^2 .
\ee
These are symmetric under interchange of $\p0''$ and $\p0'$ as a consequence of \eqref{timerev} expressing the time symmetry of the NBWF classical ensemble. 

We now {\jf make a crude} estimate of these transition probabilities in the MSSt model and deduce a few features of the resulting ensemble of bouncing histories. 
To estimate the propagator we consider the wave function on a surface at a fixed value of the scale factor $b$ near the boundary of the classical region (cf. Fig. \ref{scbound}). We consider two classical histories labeled by $\p0''$ and $\p0'$ with scalar field values $\chi''$ and $\chi'$ a distance $\Delta = \chi' - \chi'' $ apart on this constant scale factor surface.

The classical extrapolations through the bounce discussed in \cite{HHH08a} suggest the scale factor does not significantly change across the bounce. Therefore it seems a reasonable approximation to take $b$ to be constant across the quantum transition. Then the propagator defined by \eqref{propagator} is, up to multiplicative constants, that of a harmonic oscillator with mass squared $m^2b^3$ [cf. \eqref{ho}], i.e.
\be
\label{propho}
K(\chi'' ,\chi', \td) \propto \exp \left [iA(\chi''^2 + \chi'^2) -iB\chi'' \chi' \right]
\ee
with
\be
\label{propcoeff}
A\equiv\frac{\eta mb^3}{2} \cot(m\td)\ , \qquad B\equiv \frac{\eta mb^3}{\sin(m\td)}, \qquad \td \equiv t''-t ' ,
\ee
where recall $\eta\equiv3\pi/2$. 
{\jf There is, of course, a known prefactor to \eqref{propho}. But it turns out that it is the exponentials that are important for our estimate. For brevity, that's all we indicate in most formulae in this section.}

To estimate the transition amplitude \eqref{transamp} we first make the following definitions
\be
\bar\chi\equiv \frac{1}{2}(\chi''+\chi'), \qquad \Delta \equiv \chi'' - \chi'.
\label{defs1}
\ee
We then smear around the values $\chi'$ and $\chi''$ with the Gaussians
\be
\label{width} 
e^{\sigma}_{\p0''}(\chi') \propto \exp \left[-\sigma(\chi' - (\bar \chi +\Delta/2))^2\right]\ , \qquad 
e^{\sigma}_{\p0''}(\chi'') \propto  \exp \left[-\sigma(\chi'' - (\bar \chi -\Delta/2))^2\right]
\ee
where we take $\sigma$ to be sufficiently large. 
The integrals defining the transition amplitude \eqref{transamp} can be evaluated analytically. The result  is
\be
T\propto \exp\left[\frac{(-i\sigma B(\bar \chi+\Delta/2) +2\sigma (\bar \chi-\Delta/2) (\sigma -iA))^2}{4(\sigma -iA)(\sigma^2 -A^2 -2iA\sigma +B^2/4)}
-2\sigma \left(\bar \chi^2 +\frac{\Delta^2}{4}\right) +\frac{\sigma^2 (\bar \chi+\Delta/2)^2}{\sigma -iA}\right] .
\ee
If we take $\td$ to be the time a classical bounce would take, i.e. $\td=b$, then the coefficients \eqref{propcoeff} become $A \approx B/2 \approx \eta b^2/2$. In this approximation the transition probabilities \eqref{transprob} are given by
\be
p_{\rm trans}(\p0'',\p0') \propto  \exp 2\left[ \frac{(4\sigma^4 (\bar \chi-\Delta/2)^2 - 4\sigma^2 B^2\bar \chi^2)(4\sigma^3 - 2\sigma B^2) +48\sigma^5 B^2 \bar \chi (\bar \chi-\Delta/2)}{16\sigma^6 +20 \sigma^4 B^2 +4\sigma^2 B^4} \right]. \nonumber
\ee
\be
\label{transprob-ex}
\left.  -2\sigma \left(\bar \chi^2 +\frac{\Delta^2}{4}\right) +\frac{\sigma^3 (\bar \chi+\Delta/2)^2}{\sigma^2 +B^2/4}\right]
\ee
For sufficiently narrow smearing functions \eqref{width} this is approximately
\be \label{trans}
p_{\rm trans}(\p0'',\p0') \propto  \exp \left[ - \frac{ \eta^2 \Delta^2 b^4}{\sigma} \right].
\ee

As expected the quantum transition probability to a classical history with a different $\phi_0$ decreases with the difference $\Delta$ in $\chi$, or equivalently, with the difference in $\phi_0$. 

One may wonder whether our analysis also applies to the region around the bounce at maximum scale factor in recollapsing universes. As we showed in \cite{HHH08b}, recollapsing universes occur in the NBWF ensemble at low values of $\phi_0 \sim {\cal O}(1)$ if $\mu >3/2$. The classicality conditions break down near the maximum. However \eqref{trans} shows that the transition probabilities between different histories decrease 
with increasing scale factor, so we can safely conclude that the classical evolution of recollapsing universes near the turning point of the scale factor is an excellent approximation to the quantum evolution of the state.

\section{Probabilities for Bouncing Histories}
\label{probhist}

{An individual quasiclassical history in the no-boundary state can be labeled by $(\p0'',\p0')$, where $\p0'$ labels the initial (contracting) member of the NBWF classical ensemble and $\p0''$ labels the final (expanding) classical NBWF history. Both classical patches are connected by a quantum transition between them as we discussed in Section \ref{thrubounce}. We now discuss the probabilities for which of these histories occur in our universe. 

The probability that the history of  our universe has  {\jf an initial classical segment} $\p0'$ and a final classical segment $\p0''$ will be proportional to the probability $\pnb(\p0')\pnb(\p0'')$  that the classsical segments occur (cf.\eqref{nbprob}}),  multiplied by the transition probability $\ptra$ between them, viz. 
\be
\label{probhist2}
p(\p0'',{\rm trans},\p0') = \pnb(\p0'')p_{\rm trans}(\p0'',\p0')\pnb(\p0') .
\ee 
 
{\nf Consistent with the time neutrality of the NBWF}, this formula is symmetric under interchanging $\p0'$ and $\p0''$, or, put differently, with interchanging initial and final or interchanging expanding and contracting\footnote{\tf In comparing this expression with ones from elementary scattering theory the reader should keep in mind that effectively we have both initial and final quantum states.}. The ensemble of quasiclassical bouncing histories in the NBWF is time symmetric. However the individual histories {\nf need not be} time-symmetric, because the asymptotic behaviors of the expanding and contracting classical ends are different for {\nf different} $\p0$. But for every history in the ensemble the time-reversed is in the ensemble also, with equal probability. 

{\nf The time symmetry of \eqref{probhist2}} shows conclusively that there is a risk in extrapolating with classical equations through regions of configuration space where their use cannot be justified. We did that in \cite{HHH08a} and found an asymmetry between {\nf the two asymptotic regions} {\jf that does not occur in this quantum mechanical treatment.}

The probabilities \eqref{probhist2} are probabilities for the histories to occur\footnote{They are third person or bottom up probabilities in the terminology of e.g. \cite{HH15b}.}. From these various conditional probabilities may be constructed. For instance, suppose by measurements of the expansion we determine that our late time classical history is $\p0''$. What is the probability that this present history arose from a quantum transition from an earlier classical history $\p0'$? This is given by the conditional probability
\be
\label{bncpast}
p(\p0'|\p0'') \equiv p(\p0',\p0'')/p(\p0'') = \ptr(\p0'',\p0')\pnb(\p0') \ . 
\ee

Such probabilities for our past on the opposite side of a bounce are unlikely to lead to testable predictions today 
for at least two reasons:  First, as we discussed in \cite{HHH08a}, in a fully four-dimensional, history based,  formulation of quantum mechanics, we do not need to trace our history arbitrarily far back in the past to get predictions for the present. The presence or absence of a bounce or singularity prior to slow roll inflation does not affect the predictions of the CMB today. Second as, we argued in \cite{HH11a}, the physical arrows of time point away from the bounce on both sides consistent with overall time symmetry. To have an effect on our observations, signals from events on the far side would have to propagate backwards in the time direction defined by the thermodynamic arrow there and then through a quantum epoch {where the notion of a classical spacetime background is not available.}

\section{Black Holes}
\label{blackholes}

We now briefly comment on the implications of the quantum perspective on classicality that we have developed for the formation and evaporation of black holes. Our aim here is not to provide answers to well known open questions about black hole evaporation. Rather we illustrate how the questions change when the observations (a)-(d) listed in the Introduction are taken into account. 

We begin by sketching a quantum mechanical description of the process of collapse and evaporation. For simplicity we take the cosmological constant to vanish. The quantum state is then described by a wave function $\Psi$ on a configuration space of asymptotically flat three-geometries on $\bf R^3$, represented by {\jf asymptotically flat}  three-metrics $h_{ij}(\vx)$, and asymptotically vanishing configurations of a single scalar field, $\chi(\vx)$. Thus $\Psi=\Psi[\hij,\chi,t]$ where $t$ is the time at spatial infinity in a particular Lorentz frame.  The evolution of this quantum state is described {\tf in a suitable gauge} by a Wheeler-DeWitt equation (WdW) of the form
\be
i\hbar\frac{\partial \Psi}{\partial t} = {\cal H} \Psi
\label{WdW}
\ee
with $\cal H$ an operator implementation of the Hamiltonian constraint. This is of the same form as \eqref{pschrod} but not limited to minisuperspace and with a definition of the time parameter $t$ appropriate to asymptotically flat spacetimes. Any observers watching {\jf the evaporation} are described as physical systems implied by the quantum state --- not somehow outside quantum mechanics. 

Let's assume that at an initial time the wave function is peaked about a nearly flat $\hij$, a weak, pre-collapse distribution of matter field $\chi$, and nearly zero values of their conjugate momenta. Then we expect that the evolution of the wave function will  for a while remain peaked around a classical collapse history obeying the Einstein equation. Assume the initial state is such that this history eventually becomes compact enough to form trapped surfaces and, if extrapolated classically, would form a black hole. 

At late times we expect the state to predict classical spacetimes that include Hawking radiation that has been produced and remnants if any. However, in accord with point (b) we do not expect a transition to one single spacetime. Rather we expect an ensemble of classical spacetimes, all asymptotically flat, but differing at a minimum in the particular distribution of the Hawking radiation, the time the black hole exploded, and what the observers see. 
Therefore, even if the initial state is arranged to predict one initial spacetime describing a black hole collapse,  there will be an ensemble of  different  final classical spacetimes with probabilities in the final state. {\jf One can envision} a transition matrix between classical spacetimes of the kind discussed for bouncing universes  in Section \ref{thrubounce}. 

{\kf Since we have assumed asymptotic flatness its reasonable to assume that there}  are regions of classical spacetime near $\scrip$ and $\scrim$  as illustrated in Figure \ref{bhclass-sptime}. These are connected quantum mechanically  by the WdW equation. The natural question is how far into the interior do these regions of classical spacetime extend? The production of a singularity in the naive classical extrapolation of the initial configuration suggests that {\kf there will be no classical spacetime}  near the region of the classical singularity  at Planck scale matter densities. However, the case of bouncing universes discussed in this paper suggests that {\kf the limits of classical spacetime could be}  at much larger radii than those determined by arguments based on the Planck scale. In the absence of a calculation of the relevant set of coarse-grained histories one should leave both possibilities open. We now describe how a number of questions look from this point of view:

\begin{figure}[t]
\includegraphics[height=4in]{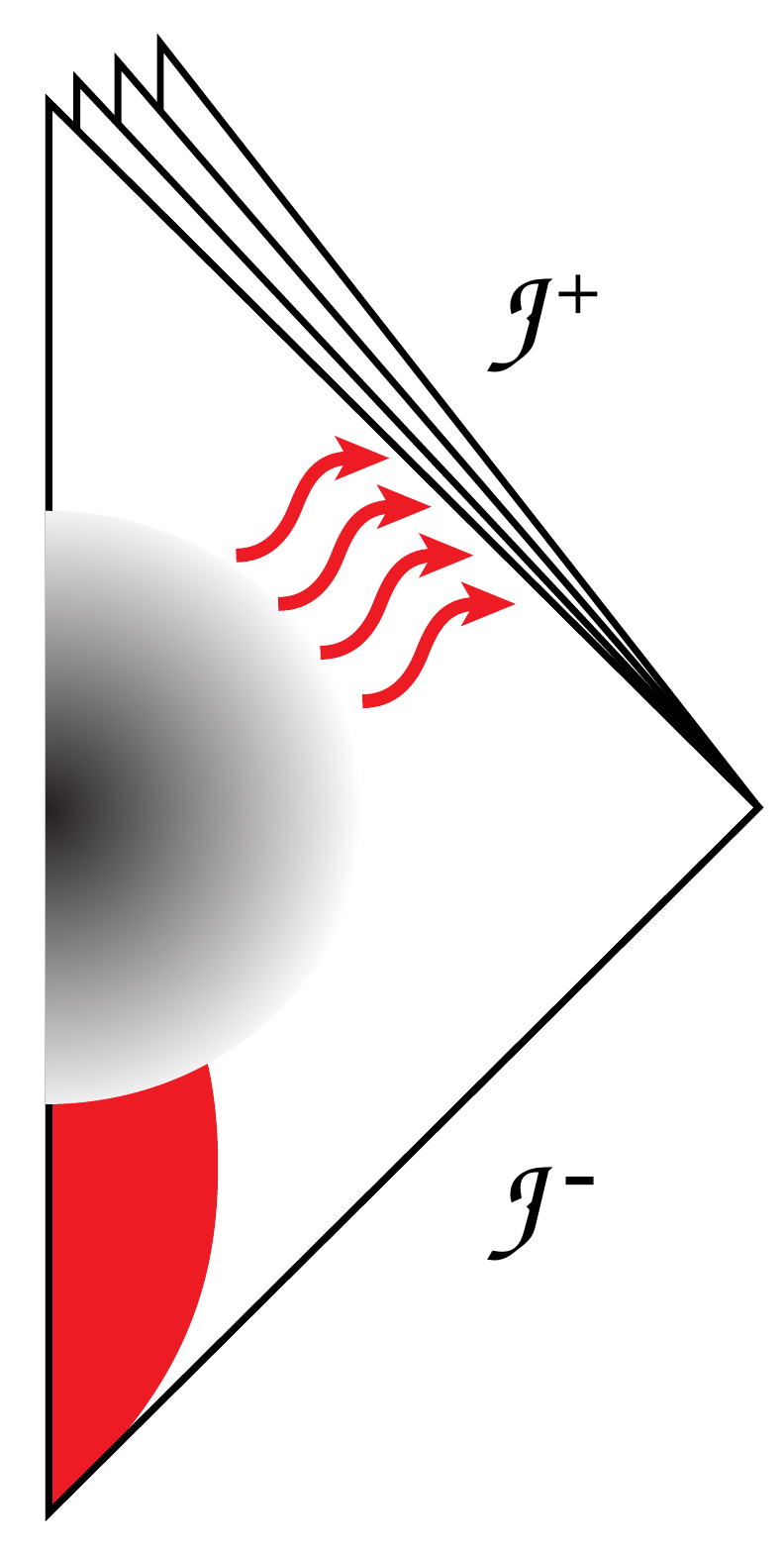} 
\caption{A schematic representation of {\jf the quantum mechanical} formation and evaporation of a black hole. The geometries describing this process are assumed to be asymptotically flat and represented here by standard conformal diagrams. Near $\scrim$ there is an initial region of classical spacetime and  matter (red) that begins to collapse classically.  In the shaded region where the classical equations would imply a singularity we expect the classical behavior of spacetime to break down. However, quantum evolution mediated by the Wheeler-DeWitt equation can still be expected to hold.  This can provide probabilities for quantum transitions of the initial classical spacetime to {\jf a set of} final classical spacetimes near $\scrip$ differing for example by the configuration of the Hawking radiation. There is no notion of an absolute event horizon as the boundary of the past of $\cal I^+$ because there is not enough classical spacetime to define it. There is then no causal obstacle to information emerging from the black hole. {\jf But the information is spread out over a range of final spacetimes.} The diagram is in many ways analogous to Figure  \ref{transition}.}
\label{bhclass-sptime}
\end{figure}

{\it What is the global structure of spacetime?:}   {\j2 There is no answer to this question since we do not expect to have a single globally defined classical spacetime. The questions are rather the following:  First: In what patches of superspace does the WdW equation plus the initial quantum state predict classical spacetime?  Second, how are histories in these patches connected quantum mechanically.  } 

{\it Is there a unitary map between the state at $\scrim$ and the state at $\scrip$?} {\j2The WdW equation \eqref{WdW} is formally unitary. In analogy with barrier penetration and bouncing universes we expect it to provide quantum probabilities for transitions between classical histories of geometry and field near $\scrim$ and classical histories of geometry and field near $\scrip$. There is no notion of unitary evolution of the state of a  field in one background spacetime because there is no one fixed spacetime. Rather one initial classical spacetime branches quantum mechanically into many final classical spacetimes\footnote{As has been noted by many, see e.g. \cite{1intomany}.}. Information in the field alone is lost because of this, but WdW evolution provides no obvious mechanism to lose total information in geometry and field.}

{\it Is information trapped behind a horizon?}  {\tf In classical general relativity the absolute event horizon is the boundary of the past of $\scrip$ {\tf of a single classical spactime.} In the quantum description of the process of collapse and evaporation sketched  here there will  be many different histories with patches of classical spacetime. There will be null surfaces in these patches (see Figure \ref{bhclass-sptime}).  But it is  possible that none of the patches will be big enough to have a region that cannot be connected by null rays to infinity. Then there will be no absolute event horizon, because there is not enough classical spacetime to construct it, and therefore no causal obstacle to complete information being available in the final state. }  

{\it Are remnants predicted?}  By remnants we mean many degrees of freedom in a small volume with low mass. The `nice slice argument'\footnote{For example as stated in \cite{Giddings07} or \cite{Pol95}} suggests that remnants will be formed if the patch of classical spacetime near $\scrim$ extends far enough into the interior to reveal an apparent horizon that reaches a sufficiently small size. The classical equations present no obstacle to such an extrapolation. But point (d) shows that this is not guaranteed. Classical spacetime can break down in low curvature regions where the classical equations of motion hold. The question becomes rather whether the quantum state predicts enough classical spacetime to make the nice slice argument. 

A way of summarizing all this is that standard arguments assuming a classical background spacetime have to be prefaced by the question: ``Does the quantum state imply sufficient classical spacetime to make this argument''. In future work we plan to return to this in more detail.

\section{Eternal Inflation}
\label{ei}

A quantum viewpoint on classicality also leads to a new perspective on the physics of eternal inflation \cite{EIrefs}.

Eternal inflation concerns a phase deep into the inflationary regime of the early universe - high up the inflaton potential and well before the usual slow roll period -  in which the quantum dynamics of fluctuations rather than the classical slow roll of the background field dominates the universeÕs evolution. It occurs in models with a regime where the inflaton potential has an extremely flat patch, relative to the value of the potential, namely when 
\be \label{EI}
V_{,\phi}^2 < V^3.
\ee
{\rf Long-wavelength fluctuations leaving the horizon when this condition holds have a large expected amplitude. This is because the variance of the Gaussian NBWF probabilities} for linear fluctuations \eqref{semiclass-flucts} around a homogeneous isotropic, classical background is essentially given by the ratio $V^3/V_{,\phi}^2$. In fact, when the condition \eqref{EI} holds the effect of the fluctuations up the potential more than compensates for the classical roll down, even though the field energies are well below the Planck scale\footnote{In a quadratic potential, for instance, the condition \eqref{EI} hold for $\phi >1/\sqrt{m}$ where $V \sim 10^{-5}$.}. This indicates that not only linearised perturbation theory breaks down in eternal inflation, but also that the common assumption that there is an approximately classical background is not justified. 

To illustrate this, a calculation in perturbation theory of the expected fractional change in the volume ${\cal V}(t)$ of a surface of constant scalar field, due to the combined effect of all fluctuations outside the horizon, yields {\uf (e.g. \cite{HHH10a})}
\be
\left \langle \frac {\delta {\cal V}}{{\cal V}_0} (t)  \right \rangle  \approx \frac{1}{8\pi^2} \frac{V^3(t)}{V_{,\phi}^2(t)} 
\label{bent}
\ee
where ${\cal V}_0(t)=2\pi^2 b^3(t)$ is the volume of a constant field surface at time $t$ in the unperturbed geometry. 
From \eqref{EI} we see that the expected volume of the reheating surface in universes with a regime of eternal inflation typically differs significantly from the reheating volume in the classical background. 

It thus appears that on the large scales associated with eternal inflation, the universe's evolution is in the first place governed by the quantum dynamics of the fluctuations and their back reaction on the geometry. The no-boundary state does {\it not} predict classical evolution, in the sense that there are no high probabilities for correlations in time given by the Einstein equation. It has been argued \cite{Star86,Linde96,Gratton05} that the quantum dynamics 
can be described by treating the fluctuations as a stochastic force term added to the classical equations. But this analysis assumes a classical background to start with. Furthermore its results merely point towards an inconsistency of this approach, since it was found that the quantum noise tends to give rise to a highly inhomogeneous - and possibly infinitely large - universe on the largest scales.

The quantum dynamics of eternal inflation, therefore, challenges the basic notion of classical evolution. The NBWF spreads in eternal inflation and has support over a broad range of very different geometries and field configurations. It is therefore implausible that it predicts {\mf any global} classical behaviour of geometry at all\footnote{See e.g. \cite{Dvali14} for a similar conclusion reached from a different perspective.}. 

Of course all statements about probabilities in quantum theory are relative to some coarse graining. One needs a certain amount of coarse graining to have well-defined probabilities at all. Similarly any conclusion about the breakdown of classical behaviour in quantum
{\rf cosmology} is relative to coarse graining. In particular one can recover classical evolution by further coarse graining: If one sums over all fluctuation modes that leave the horizon in eternal inflation one recovers an approximately homogeneous background with small fluctuations on shorter scales. The NBWF probabilities of such coarse grained backgrounds will be specified by the actions of the corresponding homogeneous saddle points and predict classical evolution. We have argued this is a reasonable approach to compute predictions for local features of the universe, such as the pattern of CMB fluctuations \cite{HHH10b}. {\uf We amplify on this in \cite{HH15c}.}

However the severe coarse graining needed to obtain classical behaviour in eternal inflation also means one loses all information about the global structure of the universe on those very large scales\footnote{This is not unlike the situation for black holes where the coarse grained classical solutions (when viewed from the outside) contain no information about the interior structure.}. This emphasizes the need for a dual description in terms of more fundamental variables, perhaps based on holography, to get a better handle on the quantum physics of eternal inflation.

\section{Discussion}
\label{conclusions}
{\j2 The conclusions of this paper are the implications  of a quantum perspective on classical evolution summarized in points (a)-(d)  listed in the Introduction, and illustrated by the various examples considered in this paper. } 

It is an inescapable inference from the physics of the last nine decades that quantum mechanics is fundamental and that classical physics is an approximation to it that emerges only in certain limited circumstances.{\j2 Classical physics holds when the probabilities predicted by a system's quantum state are high for histories exhibiting correlations in time governed by classical deterministic laws. In particular, classical spacetime is an approximation in a quantum theory of gravity holding in limited circumstances, {\nf or `patches'}, specified by the quantum state.} We should therefore not generally assume classical spacetime throughout, or classical backgrounds. Rather we should assume a quantum state and derive when and where the classical approximation holds in a background independent manner.  

It is common to assume that the classical approximation to quantum mechanics holds until the classical equations become singular or Planck scale physics is predicted by them. A lesson of this paper is that this assumption is not generally reliable. In the examples of barrier penetration, the growth of fluctuations, and bouncing universes {\nf we have seen there are quantum corrections to classical behaviour} in regions of configuration space where the classical equations remain well defined.{\j2 Black hole evaporation and eternal inflation may provide other examples where this is the case. If so, this has implications for the description of  the physics of black holes and eternal inflation.} We hope to return to these in more detail in future work.

\acknowledgments
We thank Steve Giddings, Gary Horowitz, Don Marolf, {\j2 Don Page,} Joe Polchinski, and Mark Srednicki for stimulating discussions especially about black holes. We thank Ted Jacobson for discussions and a critical reading of the paper. TH thanks the KITP and the Physics Department at UCSB for their hospitality.  JH thanks the ITP at KU Leuven for hospitality and the Wheeler family on High Island Maine where part of the paper was written. We thank Sheridan Lorenz, Chris Pope, and the Mitchell Institute for hospitality at Great Brampton House (UK). This research was  supported in part by the National Science Foundation under Grant No. PHY11-25915. The work of JH was also supported by the US NSF grant PHY12-05500. The work of TH is supported in part by the National Science Foundation of Belgium (FWO) grant G.001.12 Odysseus and by the European Research Council grant no. ERC-2013-CoG 616732 HoloQosmos. TH also acknowledges support from the Belgian Federal Science Policy Office through the Inter-University Attraction Pole P7/37 and from the European Science Foundation through the Holograv Network.

\appendix

\section{Variables and Action for the Minisuperspace Model}
\label{appa}

We use Planck units where $\hbar=c=G=1$. The minisuperspace of homogeneous, isotropic, closed three-geometries and spactial field configurations is spanned by the coordinates $x^1=b, x^2=\chi$.
The Euclidean action $I[g,\Phi]$ is a sum of a  curvature part $I_C$ and a part $I_\Phi$ for the scalar field $\Phi$.  The general form for the curvature action is: 
\begin{equation}
I_C[g] = -\frac{1}{16\pi}\int_M d^4 x (g)^{1/2}(R-2\Lambda) +\text{(surface terms)} .
\label{curvact}
\end{equation}
The general form for the matter action for a scalar field moving in a quadratic potential is:
\begin{equation}
I_{\Phi}[g,\Phi]=\frac{1}{2} \int_M d^4x (g)^{1/2}[(\nabla\Phi)^2 +m^2 \Phi^2] . 
\label{mattact}
\end{equation}
The integrals in these expressions are over the 4-disk with one boundary defining the NBWF. 
Homogeneous, isotropic, Euclidean metrics are represented as 
\be
ds^2 = 
N^2(\lambda) d\lambda^2 + a^3(\lambda) d\Omega^2_3 .
\label{homoiso-euc}
\ee
It proves convenient to also introduce rescaled measures $H$ and $\phi$ as follows: 
\begin{subequations}
\label{defs}
\begin{equation}
H^2 \equiv \Lambda/3 ,
\label{defH}
\end{equation}
\begin{equation}
\phi \equiv (4\pi/3)^{1/2} \Phi   .
\label{defphi}
\end{equation}
\end{subequations}
The scaling for $H$ is chosen so that the scale factor of a classical inflating universe is proportional to $\exp(H t)$ --- the usual definition of $H$. 
In these variables the Euclidean action takes the following simple form:
\begin{equation} 
I[a(\lambda),\phi(\lambda)] = \frac{\eta}{2}\int^1_0 d\lambda N \left\{ -a \left(\frac{a'}{N}\right)^2 -a +H^2a^3 +a^3\left[\left(\frac{\phi'}{N}\right)^2 + m^2\phi^2\right]\right\}
\label{eucact}
\end{equation}
where $\eta=3\pi/2$, $'$ denotes $d/d\lambda$, and the surface terms in \eqref{curvact} have been chosen to eliminate second derivatives.  The center of symmetry SP  and the boundary of the manifold $M$ have arbitrarily been labeled by coordinates $\lambda =0$ and $\lambda=1$ respectively. The overall  constant $\eta$ means that the action \eqref{eucact} agrees with the sum of \eqref{curvact} and \eqref{mattact} when evaluated on the metric \eqref{homoiso-euc}.

Three equations follow from extremizing the action \eqref{eucact}  with respect to $N$, $\phi$, and $a$. 
They imply the following equivalent relations:
\begin{subequations}
\label{euceqns_N}
\begin{equation}
\left(\frac{a'}{N}\right)^2 -1 +H^2a^2 +a^2\left[-\left(\frac{\phi'}{N}\right)^2 + m^2\phi^2\right]=0,
\label{eucconstraint_N}
\end{equation}
\begin{equation}
\frac{1}{a^3N}\left(a^3\frac{\phi'}{N}\right)' - m^2 \phi = 0 , 
\label{eucphieqn_N}
\end{equation}
\begin{equation}
\frac{1}{N}\left(\frac{a'}{N}\right)' +2a\left(\frac{\phi'}{N}\right)^2 + a(H^2+m^2\phi^2) = 0 \ .
\label{eucaeqn_N}
\end{equation}
\end{subequations}
These three equations are not independent. The first of them is the Hamiltonian constraint. From it, and any of the other two, the third follows. Their solutions with boundary conditions of regularity at the center of the disk and values $(b,\chi)$ on the boundary determine the complex (fuzzy) instantons whose action gives the semiclassical approximation to the NBWF in \eqref{semiclass}. 

We write metrics for homogenous, isotropic Lorentzian cosmological geometries as [cf \eqref{homoiso}]  
\be
ds^2 = -\Nh^2(\lambda) d\lambda^2 + \ah^3(\lambda) d\Omega^2_3 .
\label{homoiso1}
\ee
The action for Lorentzian histories with metrics of this form is
\begin{equation} 
{\cal  S}[\ah(\lambda),\phih(\lambda)] = \frac{\eta}{2}\int d\lambda \Nh \left\{ -\ah \left(\frac{\ah'}{\Nh}\right)^2 +\ah -H^2\ah^3 +\ah^3\left[\left(\frac{\phih'}{\Nh}\right)^2 - m^2\phih^2\right]\right\}.
\label{loract}
\end{equation}
 The resulting classical equations written in terms of of the proper time $t$ defined by  $dt=\Nh d\lambda$ are:  
\begin{subequations}
\label{loreqns}
\begin{equation}
\left(\frac{d\ah}{dt}\right)^2 +1 -H^2\ah^2 -\ah^2\left[\left(\frac{d\phih}{dt}\right)^2 + m^2\phih^2\right]=0\  ,
\label{lorconstraint}
\end{equation}
\begin{equation}
\frac{1}{\ah^3}\frac{d}{dt}\left(\ah^3\frac{d\phih}{dt}\right)+ m^2 \phi = 0\  , 
\label{lorphieqn}
\end{equation}
\begin{equation}
\frac{d^2\ah}{dt^2} +2\ah\left(\frac{d\phih}{dt}\right)^2 -\ah(H^2+m^2\phih^2) = 0 \ .
\end{equation}
\end{subequations}
These are the classical equations of motion for the model. 

\section{Parametrized Proper Time}
\label{paramproper}
This appendix reviews the procedure for promoting the proper time coordinate $t$ to the status of  a dynamical variable.
This is often called `parametrizing the time' \cite{Partime}. It can be thought of as introducing a matter degree of freedom that exactly tracks the $t$ coordinate --- an ideal clock. 

We begin with a general form for the Lorentzian action defined on a minisuperspace spanned by coodinates $x^A$:
\begin{equation}
\cS[N(\lambda),x^A(\lambda)] = \int^1_0 d\lambda \Nh  \left[\frac{1}{2} G_{AB} \left(\frac{1}{\Nh}\frac{dx^A}{d\lambda}\right) \left(\frac{1}{\Nh}\frac{dx^B}{d\lambda}\right) -\cV (x^A) \right] .
\label{mini-act}
\end{equation}
In the present case $q^A=(b,\chi)$ and $G_{AB}$ and $\cV$ can be read off \eqref{loract} as\footnote{These definitions of $G_{AB}$ and $\cV$ differ from those used in \cite{HHH08a,HH12b,HHH14} etc by incorporating factors of $\eta$ that would otherwise show up in the final equations in non intuitive ways}.
\begin{align}
\label{GV}
G_{AB}&=\eta \ {\rm diag} (-b,b^3) \nonumber \\
\cV&=\frac{\eta}{2}[-b +b^3(H^2+m^2\chi^2)]
\end{align}
$G_{AB}$ defines the DeWitt metric on minisuperspace. 

From the action \eqref{mini-act} it is straightforward to compute the momenta $p_A$ conjugate to $x^A$ and the Hamiltonian $H(p_A,x^A)$. Neglecting a multiplicative factor of $N$ the Hamiltonian is
\be
\label{Hconstr}
H(p_C,x^C) = \frac{1}{2} G^{AB}(x^C)p_Ap_B + \cV(x^C)
\ee
where $G^{AB}$ is the inverse metric to $G_{AB}$. The reparametrization invariance of the action \eqref{mini-act} implies the constraint that this Hamiltonian vanishes which is the content of \eqref{lorconstraint}. 

We now turn to promoting the proper time to the status of a dynamical variable.  Write the action \eqref{mini-act} in terms of the proper time $t$ using $Nd\lambda=dt$ and $a'/N=da/dt$. Then let $t$ be a function of a parameter $\xi$. The result is 
\be
\label{reparamact}
\cS_p[t(\xi),x^A(\xi)] = \int_0^1 d\xi \  {\dot t} \left[\frac{1}{2} G_{AB}(x^A) \left(\frac{\dot x^A}{\dot t}\right)\left(\frac{\dot x^B}{\dot t}\right)  -\cV (x^A) \right] .
\ee
where a dot means $d/d\xi$.  If the reparametrization invariance is fixed by requiring $\dot t=const.$ so that  $t=N\xi\equiv T\xi$ then this action has a much more familiar looking form  
\be
\label{rpa-fixed}
\cS_p[ x^A(t)] = \int_0^T dt  \left[\frac{1}{2} G_{AB}(x^A) \left(\frac{dx^A}{dt}\right)\left(\frac{dx^B}{dt}\right)  -\cV (x^A) \right] .
\ee

The action \eqref{reparamact} defines a reparametrization invariant theory with the dynamical variables $(t,x^A)=(t,b,\chi)$. From this it is straightforward to calculate the momenta $(p_t, p_A)$. One finds in particular
\be
\label{paraconstraint}
p_t +  H(p_A,x^A) = 0
\ee
which is the constraint arising from reparametrization invariance. 

The quantum mechanics  of this reparametrized theory is described by a wave function $\Psi=\Psi(t,x^a)$. This obeys an operator form of the constraint \eqref{paraconstraint}. To exhibit this with the factors of $\hbar$ in we momentarily revert back to $G=c=1$ units where $\hbar$ has the dimensions of length squared. The constraint is
\begin{align}
\label{seqn1}
i\hbar\frac{\partial\Psi(x^A,t)}{\partial t} =& 
H\left(-i\hbar\frac{\partial}{\partial x^A},x^A\right)\Psi(x^A,t)  \nonumber \\ 
=&\left(-\frac{1}{2} \hbar^2 \nabla^2 +\cV(x^A) \right)\Psi(x^A,t) 
\end{align}
where $\nabla^2$ is the Laplacian in the metric $G_{AB}(x^A)$.
This has the familiar form as the Schr\"odinger equation.  The explicit form of the equation is
\be
\label{seqn-explicit}
i\hbar\frac{\partial\Psi}{\partial t} = \frac{\hbar^2\eta}{2b^2}\left[\frac{\partial}{\partial b}\left(b\frac{\partial\Psi}{\partial b}\right)-\frac{1}{b}\frac{\partial^2\Psi}{\partial\chi^2}\right]+\eta\left[-b+b^3(H^2 + m^2\chi^2)\right] \Psi .
\ee
To get back to Planck units just set $\hbar=1$.

The Hamilton-Jacobi equation that arises in the semiclassical approximation for the ``energy eigenstates'' of \eqref{seqn-explicit} is found by replacing the momenta in \eqref{paraconstraint} by derivatives of the action $W$ and is
\be 
E=H\left(\frac{\partial W}{\partial x^B}, x^A\right)=\frac{1}{2}G^{AB}\frac{\partial W}{\partial x^A}\frac{\partial W}{\partial x^B} +\cV(x^A) .
\label{HJ}
\ee
Its explicit form is 
\be
E= \frac{\eta}{2}\left[-\frac{1}{b}\left(\frac{\partial W}{\partial b}\right)^2+\frac{1}{b^3}\left(\frac{\partial W}{\partial\chi}\right)^2\right]+\eta\left[-b+b^3(H^2 + m^2\chi^2)\right] .
\label{HJ-explicit}
\ee

The explicit form of the reparametization fixed action \eqref{rpa-fixed} is 
\begin{equation} 
{\cal  S}_p[\ah(t),\phih(t)] = \frac{\eta}{2}\int_0^T dt \left\{ -\ah \left(\frac{da}{dt}\right)^2 +\ah -H^2\ah^3 +\ah^3\left[\left(\frac{d\phi}{dt}\right)^2 - m^2\phih^2\right]\right\}.
\label{loract-fixed}
\end{equation}
If this is evaluated along a curve of constant $a$ we have
\be
{\cal  S}_p [a(t),\phi(t)]= c(a) + \frac{1}{2}\int_0^T  dt \  \eta a^3\left[\left(\frac{d\phi}{dt}\right)^2 - m^2\phih^2\right]
\label{ho}
\ee
where $c(a)$ is function only of $a$. This is the action for a harmonic oscillator.

\end{document}